\documentclass[11pt]{article}

\usepackage{fullpage}
\usepackage{amsmath}
\usepackage{amsfonts}
\usepackage{amssymb}
\usepackage{authblk}

\usepackage{epsfig} 
\usepackage{tikz}

\begin{document}

\title{\sf Time-delayed models of gene regulatory networks}
\author{K. Parmar, K.B. Blyuss\footnote{Corresponding author: K.Blyuss@sussex.ac.uk}, Y.N. Kyrychko}
\affil{Department of Mathematics, University of Sussex, Falmer, Brighton, BN1 9QH, UK}
\author{S.J. Hogan}
\affil{Department of Engineering Mathematics, University of Bristol, Bristol, BS8 1TR, UK}

\maketitle

\begin{abstract}
In this review we discuss different mathematical models of gene regulatory networks as relevant to the onset and development of cancer. After
discussion of alternative modelling approaches, we use a paradigmatic two-gene network to focus on the role played by time delays in the dynamics
of gene regulatory networks. We contrast the dynamics of the reduced model arising in the limit of fast mRNA dynamics with that of the full model.
The review concludes with the discussion of some open problems.
\end{abstract}

\section{Introduction}

Cancer is a complex disease, triggered by multiple mutations in various genes and exacerbated by a number of different behavioural and environmental factors. Some risk factors associated with possible onset and development of cancer are preventable, such as, inappropriate diet, physical inactivity, smoking and drinking \cite{Dan05}, while other causes include pathogens (HPV16 and HPV18 are known to cause up to 70\% of cervical cancer cases \cite{Bosh84}), as well as genetic pre-disposition. Many studies have focussed on identifying efficient genetic cancer biomarkers, such as, specific genes and groups of genes associated with significant number of cases of breast cancer \cite{Easton99}, prostate \cite{Thom08} and pancreatic cancer \cite{Wolp14}. Despite this progress, due to significant complexity associated with mutations of various cancer genes, many molecular mechanisms of oncogenesis remain poorly understood.

Recent advances in microarray and high-throughput sequencing technologies have provided pathways for measuring the expression of thousands of genes and mapping most crucial genes and groups of genes controlling different types of cancer. The networks of interactions between DNA, RNA, proteins and molecules, are defined as {\it gene regulatory networks} (GRNs). GRNs play a major role in a large number of normal life processes, including cell differentiation, metabolism, the cell cycle and signal transduction, hence, significant efforts have been made to develop mathematical techniques for their analysis \cite{Alon,Jong02,Bernot13}.

GRNs are usually formalised as networks (undirected or directed) where the nodes represent individual genes, proteins etc, and the edges correspond to some form of regulation between the nodes. In order to make progress in understanding the onset and development of cancer, as well as to develop effective drug targets, it is essential to be able to reconstruct GRNs pertinent to particular types of cancer from available data. Yeh et al. \cite{Yeh} have used a $K$-nearest-neighbours algorithm to identify GRNs correlated with cancer, tumour grade and stage in prostate cancer. As an alternative approach, Bonnet et al. \cite{Bon} have utilised LeMoNe (Learning Module Networks) algorithms to derive GRNs from gene and mRNA expression, as measured in lymphoblastoid cell lines of prostate cancer patients. A rule-based algorithm has been successfully used to determine GRNs in colon cancer \cite{Wang10}, and similar kinds of networks have been identified from microarray data using neural fuzzy networks \cite{Vin12}. Madhamshettiwar et al. \cite{Mad12} discuss different approaches to infer GRNs in ovarian cancer, as well as the potential of using these GRNs to develop optimal drug targets. Bayesian network techniques have been employed to construct GRNs from microarray data for breast cancer \cite{Akh12}. In a recent paper, Emmert-Streib et al. \cite{ES14} have successfully used a BC3Net inference algorithm to analyse a large-scale breast cancer gene expression data set and reconstruct the associated GRN.

In the next section we survey and compare different approaches to model the dynamics of GRNs, making an emphasis on particular biological features that can be best represented by each of the methods. Section 3 focuses on the role of transcriptional and translational time delays in GRN models, and we show how such delays can be introduced in a paradigmatic two-gene activator-inhibitor GRN. Depending on a particular biological regime in which a given GRN is operating, it is often possible to encounter a situation where there is a significant separation of time scales due to, for instance, very fast mRNA dynamics compared to other characteristic time scales. In such a case it is possible to perform dimensional reduction and concentrate on the dynamics of a smaller number of variables. In Section 4 we analyse such reduced model and show how one can derive analytical conditions that lead to a transition from a stable steady state to stable periodic oscillations that are impossible in the model system without the time delays. Section 5 extends the analysis to a full nonlinear system to illustrate differences in stability conditions. The review concludes in Section 6 with a discussion of main results and some open questions.

\section{Mathematical models of gene regulatory networks}

In the analysis of gene regulatory networks and their dynamics, the first step is the identification of key modules or components and possible relations between them, which is often done by interrogating available expression data. Once the topology of the GRN has been fixed, the next step in modelling the dynamics is making realistic assumptions about specific rules that govern the expression of particular genes. Depending on the level of understanding of underlying processes, the complexity of the GRN under investigation, and the specific questions to be addressed, there are several methodologically different approaches that can be employed. Endy \& Brent \cite{Endy} and Hasty et al. \cite{hasty} discuss biological underpinnings for studying and modelling GRNs, while excellent reviews by de Jong \cite{Jong02}, Bernot et al. \cite{Bernot13}, Tu\v{s}ek \& Kurtanjek \cite{Tusek}, and Hecker et al. \cite{Hecker} give an overview of mathematical and statistical techniques that have been successfully used to model GRNs, and some of these methods are discussed below.

\subsection{Boolean networks}

Some of the first models developed for modelling GRNs were the so-called {\it Boolean networks} \cite{Kau1,Thom73,Som96}, where the states of all genes participating in the interactions are represented by binary variables having the values of ON and OFF, or 1 and 0, with the possibility of either synchronous or asynchronous update rules for the nodes. Boolean logic rules are then used to approximate regulatory control of gene expression \cite{Albert}, with updates of binary states of all genes taking place simultaneously \cite{Smo}. Boolean networks approach has been extended in several directions to provide a better approximation of real GRNs. Shmulevich et al. \cite{Shm} have proposed a probabilistic analogue of Boolean networks to account for stochastic nature of many processes involved in gene expression. Silvescu and Honavar \cite{SH01} have proposed temporal Boolean networks, where the next state of genes in the networks is determined not only by their current state, but also by a fixed number of their previous states, which effectively allows one to take into account some history of transitions in a GRN. Recently, Boolean network models of GRNs have been compared to models based on ordinary differential equations (ODEs), and, in fact, it has been shown that some Boolean models can be rigorously derived as coarse-grained analogues of some ODE models \cite{David08}.

Significant advantage of using Boolean networks to model GRNs lies in the fact that they allow one to consider networks with a very large number of nodes. At the same time, there are several deficiencies in this approach. The first one concerns the fact that since the gene states only admit the values of ON or OFF, this formalism does not take into account intermediate stages of gene expression \cite{Xu07}. Another issue is that GRNs modelled by Boolean networks can exhibit behaviour not observed in real life, hence, special care has to be taken when choosing the class of admissible Boolean functions \cite{Karlson}.

\subsection{Fuzzy methods}

Due to intrinsic imprecision and uncertainty associated with gene expression data, it may be appropriate to move away from precise rules of Boolean logic in favour of machine learning techniques based on fuzzy logic. The basic idea is that rather than trying to reconstruct some assumed fixed gene network topology, one considers the whole family of possible networks with all possible distributions of links between nodes. The problem lies in using actual data to assign appropriate probabilities to each of these configurations, so that for a given input the fuzzy network would provide an output that most resembles actual data. A significant advantage of fuzzy logic for inferring the structure of GRNs lies in their ability to rely on already available knowledge of biological relations between different nodes in the network, and, at the same time, being able to recover important previously unknown connections. On the other hand, fuzzy methods for GRN inference are characterised by a high level of computational complexity.

To give a few examples,  fuzzy approach has been used to analyse microarray data from the yeast cell cycle and to recover a set of GRNs, with $k$-nearest-neighbour algorithm being used to replace missing data \cite{Brock09}. Woolf and Wang \cite{Woolf} have used a $k$-means clustering algorithm to reconstruct and evaluate GRNs for {\it Saccharomyces cerevisiae}. In this approach, groups of co-regulated genes are considered as clusters, and clustering algorithm is then used to detect cluster centres. Volkert and Mahlis \cite{Volkert} have used a smooth response surface algorithm to recover GRNs from gene expression data for {\it Saccharomyces cerevisiae}. Approaches based on an artificial bee colony search algorithm have allowed the reconstruction of a GRN in {\it Escherichia coli} \cite{Das09}. A very recent review by Al Qazlan et al. \cite{Qaz} gives an overview of different fuzzy methods, as well as their combinations with other approaches, such as ordinary differential equations, with the purpose of optimising data mining of gene expression and microarray datasets to recover GRNs.

\subsection{Ordinary and delay differential equation models}

A very powerful and mathematically insightful methodology for analysis of GRNs is based on nonlinear ordinary or delay differential equations (ODEs or DDEs). In this approach, a gene regulatory network is represented by concentrations of different mRNAs and proteins, and the dynamics can be written as a system of ODEs or DDEs using the law of mass action for individual reactions \cite{Alon}. Some of the earliest results on ODE models of gene regulation go back to Goodwin \cite{Good63,Good65}, who introduced and studied a negative feedback loop involving the concentrations of mRNA, an enzyme and a metabolite. It has been later shown that a negative feedback loop is absolutely essential to ensure the existence of stable periodic solutions, while positive feedback is required for multi-stationarity \cite{Gou98,Sno98}. This approach was subsequently generalised and expanded \cite{TO78,Kel94,HS96,CA00}; reviews by Smolen et al. \cite{Smo}, de Jong \cite{Jong02} and Hecker et al. \cite{Hecker} discuss some of these models based on systems of nonlinear ODEs. A very important aspect of all these models is a {\it regulation function} that controls the rates of gene expression. In light of experimental evidence suggesting monotonic sigmoidal shape of regulation functions \cite{yag75}, a conventional choice for this function is given by the Hill function \cite{polynikis09,widder07,hill2}. Weiss \cite{hill1} has discussed various chemical mechanisms associated with the Hill function, including different kinds of ligand binding, and a more recent review of the uses of the Hill function in GRN models can be found in \cite{sant}.

In order to more accurately represent a switch-like behaviour of the gene expression, several authors have developed models of GRNs using piecewise-linear differential equations, in which the continuous Hill function is replaced by a discontinuous step function \cite{gla73,glass75,Plah98,plah05,casey06,Geb07}.
Besides regular steady states, the piecewise-linear models also allow for singular steady states, which although important for representing homeostasis in GRNs, are complex to analyse due to discontinuities at the thresholds \cite{Plah94,SnoTho}. Polynikis et al. \cite{polynikis09} discuss various features of piecewise-linear ODE models and different dynamical regimes that can be exhibited in these models, including possible periodic solutions, sharp-threshold dynamics, and the comparison with models based on continuous regulation function.

In terms of applications to cancer, ODE models have explained aberrant dynamics of the NF-$\kappa$B transcription factor linked to oncogenesis, tumour progression and resistance to therapy, as well as the dynamics of I$\kappa$B-NF-$\kappa$B \cite{hoff02,werner05}. Another example is the analysis of the feedback loop between the tumour suppressor p53 and the oncogene Mdm2 \cite{geva06}, and the single-cell response of p53 to radiation-induced DNA damage \cite{ma_pnas05}. There is a clinical evidence suggesting that different components of the PI3K/AKT pathway can lead to aberrant cell growth,
metastatic competence and therapy resistance, and some progress has been made in modelling this pathway and identifying inhibitors responsible for the regulation of PI3K/AKT signalling \cite{hen05}. Cheng {\it et al.} \cite{cheng} and Edelman {\it et al.} \cite{edelman10} give a number of examples of the uses of differential equation based models for the analysis of GRNs in cancer.

Another aspect that has to be properly accounted for in dynamical models is the fact that transcription and translation during gene expression often take place over non-negligible time periods. Monk \cite{monk03} has shown how time delays can cause oscillatory gene expression and provide insights into the dynamics of interactions between p53 and Mdm2 proteins associated with cancer suppression. Subsequent research has focused on the role of time delays in GRN dynamics \cite{wan09,wang10,bodnar11,wang12,xiao14}. Xiao and Cao \cite{xiao08} have analysed a Hopf bifurcation in a gene network with two transcriptional delays, which occurs when the sum of the delays passes through a critical value, and shown how the amplitude and period of oscillations of gene expression change with the time delays. Due to the fact that it may not be practically possible to identify discrete transcription/translation time delays, a better alternative would be to use models with distributed delay \cite{HeCao}. Models with time delays have been used to understand the regulation of feedback loops involving transcription factors E2F and Myc, known oncogenes and possible tumour suppressors \cite{aguda08,aguda13}. Ribeiro et al. \cite{RibGreKau} have developed a delayed stochastic simulation algorithm for analysis of the p53-Mdm2 feedback loop whose malfunction is associated with 50\% of cancers. Sequences of multiple reactions with unknown intermediate kinetics can also be successfully analysed using time-delayed models \cite{macd77,IL91}.

\subsection{Stochastic models}

Experimental evidence suggests that significant stochastic fluctuations are observed during gene expression and regulation, hence, in many cases it is paramount to use stochastic models for studying GRN dynamics \cite{Pau05,Ros07}. Even in the absence of extrinsic noise associated with variability in different environmental factors, there are several fundamental processes responsible for intrinsic stochasticity of gene expression \cite{Raj,Swain}. One of these is the process of initiation of transcription, which starts by first forming an {\it elongation complex} by binding RNA polymerase (RNAp) to the promoter region of the gene, and there is a significant variation in the duration of elongation processes between different transcription events \cite{herbert,ribeiro09,voliotis,ribeiro10}. Binding of RNAp to the promoter regions of different genes results in switching of these genes on and off, thus either blocking or facilitating further transcription, which gives another major source of noise in GRNs. Stochasticity in expression of individual gene results in stochastic behaviour of larger genetic circuits and GRNs \cite{Pau05,dunlop08}. Some of the early work on stochastic gene expression emerged from experiments in synthetic biology \cite{elo00,elo02} that demonstrated how stochasticity can result in sustained oscillations, and significant amount of research has been subsequently done both theoretically and experimentally on the analysis of stochastic (and delayed) oscillations in gene regulatory networks \cite{Swain,rosen06,kaern,Bra05}. Zavala and Marquez-Lago have recently considered delay-induced oscillations in deterministic and stochastic models of single-cell gene expression, highlighting important differences between these two types of models and associated behaviours \cite{zavala}.

Besides being an intrinsic feature of biological dynamics, stochasticity has proved to be important in the context of engineered genetic switches \cite{elo00,Gard00}. de Jong \cite{Jong02} and El Samad et al. \cite{ElSam05} discuss various methods for modelling stochastic GRN models, including stochastic master equation and various stochastic simulation algorithms. Bratsun et al. \cite{Bra05} have developed an algorithm for analysis of non-Markovian dynamics in GRNs with time delays and showed that these delays are able to induce oscillatory dynamics in the case where deterministic models do not exhibit oscillations. This methodology was later improved, and several exact stochastic simulation algorithms have been developed for simulations of time-delayed models \cite{barrio06,cai07}. A review by Ribeiro \cite{ribeiro10} discusses various techniques for simulating stochastic time-delayed dynamics of gene expression, and very recently Jansen et al. \cite{jansen15} have reviewed the role of delay distribution in the stochastic dynamics during gene expression.

Another way to approach stochasticity in the analysis and reconstruction of GRNs is by using so-called {\it Bayesian networks} \cite{Fried00}, where gene expression values are represented as random variables, and relations between them are probabilistic. Learning techniques for Bayesian networks \cite{Heck,Wer07} allow one to combine expression data with an {\it a priori} knowledge to deduce the structure of GRN that best matches the available expression data. Friedman et al. \cite{Fried00} have developed an algorithm for deriving Bayesian networks that circumvents a dimensionality problem, and this method has been used to analyse the cell cycle data for {\it S. cerevisiae}  containing numerous measurements of mRNA expression levels \cite{Spell}. Out of 800 genes it was  possible to identify a few genes controlling the regulation of cell cycle processes.

The rest of this paper is devoted to consideration of the effects of transcriptional and translational time delays on the dynamics of GRNs. In the next Section we introduce the time-delayed model of a two-gene activation-inhibition network together with its quasi-steady state simplification, and establish the well-posedness of both models. Section 4 contains the derivation of analytical conditions for stability and a Hopf bifurcation in the case of very fast mRNA dynamics, while in Section 5 the analysis is extended to the full time-delayed system. The paper concludes in Section 6 with discussion of results and future research directions.

\section{Time-delayed models: derivation and positivity}

To motivate the analysis of time-delayed effects in gene regulatory dynamics, following Polynikis et al. \cite{polynikis09}, we consider an activation-inhibition two-gene GRN consisting of two genes $a$ and $b$, which are assumed to have no effect on their own expression; at the same time, protein $P_b$ is assumed to activate the expression of gene $a$, while protein $P_a$ inhibits the expression of gene $b$. This is one of the fundamental motifs, which has been shown to be functionally relevant in GRNs \cite{cheng,prill}. Denoting the concentrations of proteins $P_a$ and $P_b$ as $p_a$ and $p_b$, and concentrations of transcribed mRNAs as $r_a$ and $r_b$, the following system of equations can be derived for the dynamics of this GRN \cite{polynikis09}
\begin{equation}\label{eqn:CNM}
\begin{array}{l}
\dot{r}_a = m_ah^+(p_b;\theta_b,n_b)-\gamma_a r_a,\\\\
\dot{r}_b = m_bh^-(p_a;\theta_a,n_a)-\gamma_br_b, \\\\
\dot{p}_a = k_ar_a-\delta_a p_a,\\\\
\dot{p}_b= k_br_b-\delta_bp_b, 
\end{array}
\end{equation}
where $m_i$ are the maximum transcription rates, $k_i$ are the translation rates, $\gamma_i$ are the mRNA degradation rates, and $\delta_i$ are the protein degradation rates for $i=a,b$. Equations (\ref{eqn:CNM}) are called the complete nonlinear model (CNM). To make further analytical progress, the activation and inhibition functions in the system (\ref{eqn:CNM}) can be written as the following Hill functions,
\[
\begin{array}{l}
\displaystyle{h^+(p_i;\theta_i,n_i)=\frac{p_i^{n_i}}{p_i^{n_i}+\theta_i^{n_i}},\mbox{ and }}\\\\
\displaystyle{h^-(p_i;\theta_i,n_i)=1-h^+(p_i;\theta_i,n_i)=\frac{\theta_i^{n_i}}{p_i^{n_i}+\theta_i^{n_i}},\hspace{0.5cm}i=a,b,}
\end{array}
\]
where $\theta_a$ and $\theta_b$ are known as activation and inhibition coefficients, and the integer parameters $n_{a}$ and $n_b$, known as Hill coefficients, determine the steepness of Hill curves \cite{Alon}. The parameters $\theta_a$ and $\theta_b$ give the values of protein concentrations $p_a$ and $p_b$, at which the corresponding Hill function achieves half of its maximum value. Depending on the values of transcription rates, this would then lead to a significant increase in the respective mRNAs regulated by these proteins \cite{Bernot13,polynikis09}.

Due to the fact that the dynamics of mRNA is normally much faster that that of related proteins, one can use a quasi-steady state assumption to simplify the CNM (\ref{eqn:CNM}) by reducing the number of equations. Effectively, this means assuming that mRNAs have already reached their steady-state concentrations, i.e. taking $\dot{r}_i\approx 0$, $i=a,b$ in the CNM (\ref{eqn:CNM}), and then focusing on the dynamics of proteins only, as given by the following simplified nonlinear model (SNM)
\[
\begin{array}{l}
\dot{p}_a = k'_a h^+(p_b;\theta_b,n_b)-\delta_ap_a,\\\\
\dot{p}_b = k'_bh^-(p_a;\theta_a,n_a)-\delta_bp_b,
\end{array}
\]
where
\begin{equation}\label{kab}
\displaystyle{k'_a=\frac{m_ak_a}{\gamma_a},\hspace{0.5cm}k'_b=\frac{m_bk_b}{\gamma_b}.}
\end{equation}
Polynikis et al. \cite{polynikis09} have shown that, while the CNM exhibits Hopf bifurcation of a positive equilibrium, leading to persistent oscillations, in the case of the SNM model this behaviour can disappear. They have also demonstrated an important role played by the Hill coefficients, as well as the separation of timescales between mRNA and proteins, with a larger scale separation favouring a stable equilibrium rather than oscillatory behaviour.

While the transcription and translation may be faster than characteristic times associated with significant changes in protein concentrations (of the order of 5 minutes for transcription+translation and 1 hour for a 50\% change in the concentration of translated protein for {\it E. coli.} \cite{Alon}), these are, in fact, multi-step processes consisting of thousands of consecutive chemical reactions. Hence, the duration of transcription and translation is non-negligible when considered in the context of GRN dynamics \cite{ribeiro10,jansen15}, and has to be correctly accounted for in mathematical models. To analyse the effects of transcriptional and translational time delays we introduce the following model
\begin{equation} \label{eqn:DCNM}
\begin{array}{l}
\dot{r}_a= m_ah^+(p_b(t-\tau_{r_a});\theta_b,n_b)-\gamma_a r_a,\\\\
\dot{r}_b= m_bh^-(p_a(t-\tau_{r_b});\theta_a,n_a)-\gamma_br_b,\\\\
\dot{p}_a= k_ar_a(t-\tau_{p_a})-\delta_ap_a,\\\\
\dot{p}_b= k_br_b(t-\tau_{p_b})-\delta_bp_b,
\end{array}
\end{equation}
where $\tau_{r_a}$ and $\tau_{r_b}$ are the delays during transcription of mRNAs $r_a$ and $r_b$, and $\tau_{p_a}$ and $\tau_{p_b}$ are the delays during translation of proteins $p_a$ and $p_b$, respectively. This model will be referred to as the delayed complete non-linear model (DCNM). Similar to the case of instantaneous transcription and translation, the quasi-steady-state assumption simplifies the system (\ref{eqn:DCNM}) to the following delayed simplified non-linear model (DSNM)
\begin{equation} \label{eq:DSNM}
\begin{array}{l}
\dot{p}_a = k'_a h^+(p_b(t-\tau_{r_a}-\tau_{p_a});\theta_b,n_b)-\delta_ap_a, \\\\
\dot{p}_b = k'_bh^-(p_a(t-\tau_{r_b}-\tau_{p_b});\theta_a,n_a)-\delta_bp_b,
\end{array}
\end{equation}
with parameters $k'_a$ and $k'_b$ defined in (\ref{kab}).

Before proceeding with the analysis, one has to augment the models (\ref{eqn:DCNM}) and (\ref{eq:DSNM}) with the appropriate initial conditions and establish that these models are well-posed, i.e. that their solutions remain non-negative for all time to ensure their biological feasibility. The initial conditions for the DCNM model (\ref{eqn:DCNM}) are given by
\begin{equation}\label{ICs}
\begin{array}{l}
r_a(s)=\phi_1(s),\hspace{0.5cm}s\in [-\tau_{\rm max},0],\\\\
r_b(s)=\phi_2(s),\hspace{0.5cm}s\in [-\tau_{\rm max},0],\\\\
p_a(s)=\phi_3(s),\hspace{0.5cm}s\in [-\tau_{\rm max},0],\\\\
p_b(s)=\phi_4(s),\hspace{0.5cm}s\in [-\tau_{\rm max},0],\\\\
\end{array}
\end{equation}
where $\tau_{\rm max}=\max(\tau_{r_a},\tau_{r_b},\tau_{p_a},\tau_{p_b})$ and $\phi_i(s)\in C([-\tau_{\rm max},0],\mathbb{R})$ with $\phi_i(s)\geq 0$ $(-\tau_{\rm max}\leq s\leq 0$, $i=1,..,4)$, and similarly for the DSNM model (\ref{eq:DSNM}). Here, $C([-\tau_{\rm max},0],\mathbb{R})$ is the Banach space of continuous mappings of the interval $[-\tau_{\rm max},0]$ onto $\mathbb{R}$. It is further assumed that $r_a(0)>0$, $r_b(0)>0$ to ensure that at least some amount of proteins will be produced.

We now prove that the solution $(r_a(t),r_b(t),p_a(t),p_b(t))$ of the DCNM model (\ref{eqn:DCNM}) with the initial condition (\ref{ICs}) is positive for all $t> 0$. This result can be proven by contradiction, following the methodology used in \cite{kyrychko05}. As a first step, let us show that $r_b(t)\geq 0$ for all $t>0$. Let $t_1>0$ be the first time when $p_a(t_1)r_b(t_1)=0$; assuming that $r_b(t_1)=0$ implies $p_a(t)\geq 0$ for all $t\in[0;t_1]$, and since $t_1$ is the first time when $r_b(t_1)=0$, this also means $dr_b(t_1)/dt\leq 0$, i.e. the function $r_b(t)$ is decreasing at $t=t_1$. On the other hand, evaluating the second equation of the system (\ref{eqn:DCNM}) at $t=t_1$ yields
\[
\frac{dr_b(t_1)}{dt}=\frac{m_b\theta_a^{n_a}}{p_a(t_1-\tau_{r_b})^{n_a}+\theta_a^{n_a}}>0,
\]
which gives a contradiction. Since $r_b(0)>0$, this implies $r_b(t)> 0$ for all $t>0$. Now that the positivity of $r_b(t)$ has been established, let $t_2>0$ be the first time when $p_b(t_2)=0$. In order for this to happen, one must have $dp_b(t_2)/dt\leq0$, i.e. the function $p_b(t)$ should be decreasing at $t=t_2$. At the same time, evaluating the last equation of the system (\ref{eqn:DCNM}) at $t=t_2$ yields
\[
\frac{dp_b(t_2)}{dt}=k_br_b(t_2-\tau_{p_b})>0,
\]
which gives a contradiction, and, therefore, $p_b(t)>0$ for all $t>0$. In a similar manner, the positivity of $p_b(t)$ implies the positivity of $r_a(t)$, which in turn implies the positivity of $p_a(t)$. Hence, all solutions $r_a(t)$, $r_b(t)$, $p_a(t)$ and $p_b(t)$ of the DCNM model (\ref{eqn:DCNM}) are positive for all $t>0$. The same approach can be employed to show positivity of solutions of the DSNM model (\ref{eq:DSNM}).

Steady states $(\bar{r}_a,\bar{r}_b,\bar{p}_a,\bar{p}_b)$ of the DCNM model can be found as roots of the following system of algebraic equations
\[
\begin{array}{l}
 m_ah^+(\bar{p}_b;\theta_b,n_b)-\gamma_a \bar{r}_a = 0,\\\\
 m_bh^-(\bar{p}_a;\theta_a,n_a)-\gamma_b\bar{r}_b = 0,\\\\
 k_a\bar{r}_a-\delta_a\bar{p}_a= 0,\\\\
 k_b\bar{r}_b-\delta_b\bar{p}_b= 0.
 \end{array}
\]
This gives
\[
\displaystyle{\bar{r}_a= \frac{\delta_a}{k_a}\bar{p}_a,\hspace{0.5cm}
 \bar{r}_b= \frac{\delta_b}{k_b}\bar{p}_b,\hspace{0.5cm}\bar{p}_b = \frac{\phi_b\theta^{n_a}_a}{\theta^{n_a}_a+\bar{p}^{n_a}_a},}
\]
where $\bar{p}_a$ satisfies the polynomial equation
\begin{equation}\label{paeq}
\displaystyle{\theta^{n_b}_b\sum_{k=0}^{n_b}\binom{n_b}{k}\bar{p}^{n_a(n_b-k)+1}_a\theta^{n_ak}_a+(\bar{p}_a-\phi_a)(\phi_b\theta^{n_a}_a)^{n_b}=0,}
\end{equation}
and we used the notation
\[
\displaystyle{\phi_a=\frac{m_ak_a}{\gamma_a\delta_a},\hspace{0.5cm}\phi_b = \frac{m_bk_b}{\gamma_b\delta_b}.}
\]
Even for realistically small values of Hill coefficients, such as $n=2,3$ \cite{Tied12} or $n=4$-$8$ \cite{Li13}, the equation (\ref{paeq}) is too complicated to allow one to analytically find closed form expressions for $\bar{p}_a$ and other state variables. Despite not having explicit formulae for possible steady states $(\bar{r}_a,\bar{r}_b,\bar{p}_a,\bar{p}_b)$, one can still perform the analysis of stability in terms of system parameters, and such results would be valid for the values of steady state variables that can be accurately and efficiently determined through numerical solution of the polynomial equation (\ref{paeq}).

\section{Analysis of the delayed simplified non-linear model (DSNM)}

In order to gain some first insights into the role of transcriptional and translational delays on the dynamics of GRN, we focus on the behaviour of the delayed simplified non-linear model (DSNM) (\ref{eq:DSNM}). To reduce the number of free parameters in the model, we introduce the new variables
\begin{equation}\label{newvar}
\hat{p}_a(t)=p_a(t),\quad \hat{p}_b(t)=p_b(t-\tau_{r_a}-\tau_{p_a}),
\end{equation}
which transform the first equation of the system (\ref{eq:DSNM}) into
\[
\dot{p}_a = k'_a h^+(p_b(t-\tau_{r_a}-\tau_{p_a});\theta_b,n_b)-\delta_ap_a\quad\Longleftrightarrow
\quad\dot{\hat{p}}_a(t)=k'_ah^+(\hat{p}_b(t);\theta_b,n_b)-\delta_a\hat{p}_a(t). 
\]
The second equation of the system (\ref{eq:DSNM}) evaluated at $t-\tau_{r_a}-\tau_{p_a}$ has the form
\[
\dot{p}_b(t-\tau_{r_a}-\tau_{p_a})= k'_bh^-(p_a(t-\tau_{r_a}-\tau_{p_a}-\tau_{r_b}-\tau_{p_b});\theta_a,n_a)-\delta_bp_b(t-\tau_{r_a}-\tau_{p_a}),
\]
and in terms of the new variables (\ref{newvar}) this can be rewritten as
\[
\dot{\hat{p}}_b(t)=k'_bh^-(\hat{p}_a(t-\tau_{r_a}-\tau_{p_a}-\tau_{r_b}-\tau_{p_b});\theta_a,n_a)-\delta_b\hat{p}_b(t).
\]
Thus, the system (\ref{eq:DSNM}) takes a form
\begin{equation}\label{pab_new}
\begin{array}{l}
\dot{\hat{p}}_a(t)=k'_ah^+(\hat{p}_b(t);\theta_b,n_b)-\delta_a\hat{p}_a(t),\\\\
\dot{\hat{p}}_b(t)=k'_bh^-(\hat{p}_a(t-\tau);\theta_a,n_a)-\delta_b\hat{p}_b(t),
\end{array}
\end{equation}
where
\[
\tau=\tau_{r_a}+\tau_{p_a}+\tau_{r_b}+\tau_{p_b},
\]
is the new combined time delay. The equation for characteristic eigenvalues $\lambda$ of the linearisation near a steady state $(\bar{p}_a,\bar{p}_b)$ of the system (\ref{pab_new}) has the form
\begin{equation}\label{eqn:char_DSNM}
 (\lambda+\delta_a)(\lambda+\delta_b)+D_{\text{DSNM}}e^{-\lambda\tau}=0, 
\end{equation}
where
\[
\displaystyle{D_{\text{DSNM}}=k'_ak'_bn_an_b\frac{\theta^{n_a}_a\theta^{n_b}_b\bar{p}^{(n_a-1)}_a\bar{p}^{(n_b-1)}_b}{(\theta^{n_a}_a+\bar{p}^{n_a}_a)^2(\theta^{n_b}_b+\bar{p}^{n_b}_b)^2}=
n_an_b\delta_a\delta_b\frac{\bar{p}^{n_a}_a}{\theta^{n_a}_a+\bar{p}^{n_a}_a}\frac{\theta^{n_b}_b}{\theta^{n_b}_b+\bar{p}^{n_b}_b}.}
\]
\noindent In the limit $\tau=0$, this equation reduces to the quadratic equation \cite{polynikis09}
\begin{equation}\label{eq:char_DSNM_tau0}
 \lambda^2+(\delta_a+\delta_b)\lambda+\delta_a\delta_b+D_{\text{DSNM}}=0,
\end{equation}
whose roots always have negative real parts, since $\delta_a>0$, $\delta_b>0$ and $D_{\text{DSNM}}>0$. This implies that for $\tau=0$, the steady state $(\bar{p}_a,\bar{p}_b)$ is stable for any values of parameters. To investigate whether this steady state can lose stability for $\tau>0$, one can note that $\lambda=0$ is not a solution of the characteristic equation (\ref{eqn:char_DSNM}). Hence, the only possible way how the steady state $(\bar{p}_a,\bar{p}_b)$ can lose its stability is when a pair of complex conjugate eigenvalues crosses the imaginary axis. In the light of this observation, one can look for eigenvalues of the equation (\ref{eqn:char_DSNM}) in the form $\lambda=i\omega$ for some real $\omega>0$. Substituting this into Eq.~(\ref{eqn:char_DSNM}) and separating into real and imaginary parts gives
\begin{equation} \label{eq:sin_DSNM}
\begin{array}{l}
 \omega^2-\delta_a\delta_b=D_{\text{DSNM}}\cos(\omega\tau),\\\\
 (\delta_a+\delta_b)\omega=D_{\text{DSNM}}\sin(\omega\tau).
\end{array}
\end{equation}
Squaring and adding these two equations yields the following equation for $z=\omega^2$
\begin{equation}
h(z)=z^2+(\delta_a^2+\delta_b^2)z+\delta_a^2\delta_b^2-D_{\text{DSNM}}^2=0,
\end{equation}
which can be solved to give the critical frequency as
\begin{equation}\label{om_dsnm}
\displaystyle{\omega_0^2=\frac{1}{2}\left[-(\delta_a^2+\delta_b^2)+\sqrt{(\delta_a^2+\delta_b^2)^2-4(\delta_a^2\delta_b^2-D_{\text{DSNM}}^2)}\right]}.
\end{equation}

One should note that $\omega_0^2$ will only admit real values, provided $\delta_a\delta_b<D_{\text{DSNM}}$, which implies that 
for $\delta_a\delta_b\geq D_{\text{DSNM}}$, the steady state $(\bar{p}_a,\bar{p}_b)$ is stable for all values of the time delay $\tau$. Note that
\begin{equation}\label{eq:h'(z)_DSNM}
\frac{dh(z)}{dz}=2z+\delta_a^2+\delta_b^2>0 \hspace{3mm} \mbox{for any } \hspace{2mm} z\geq0.
\end{equation}

The critical value of the time delay $\tau$ can be found from (\ref{eq:sin_DSNM}), which gives
\begin{equation}
 \tau_{0,n}=\frac{1}{\omega_0}\left[\mbox{Arctan}\left(\frac{(\delta_a+\delta_b)\omega_0}{\omega_0^2-\delta_a\delta_b}\right)+n\pi\right],\hspace{0.5cm}n=0,1,2,...,
\end{equation}
where $\omega_0$ is determined by Eq.~(\ref{om_dsnm}), and Arctan corresponds to the principal value of arctan. When $\tau=\tau_{0,n}$, the characteristic equation (\ref{eqn:char_DSNM}) has a pair of purely imaginary roots. To determine, whether or not these roots do indeed cross the imaginary axis, we consider $\lambda(\tau)=\mu(\tau)+i\omega(\tau)$ as a root of the equation (\ref{eqn:char_DSNM}) near $\tau=\tau_{0,n}$, satisfying $\mu(\tau_{0,n})=0$, $\omega(\tau_{0,n})=\omega_0$, $j=0,1,2,...$. Substituting $\lambda=\lambda(\tau)$ into the equation (\ref{eqn:char_DSNM}) and differentiating with respect to $\tau$ yields
\[
\displaystyle{\left(\frac{d\lambda}{d\tau}\right)^{-1}=\frac{(2\lambda+\delta_a+\delta_b)e^{\lambda\tau}}{\lambda D_{\text{DSNM}}}-\frac{\tau}{\lambda}.}
\]

\noindent From this equation, one can find
\[
\begin{array}{l}
\displaystyle{\mbox{sgn}\left\{ \left[\frac{d(Re\lambda)}{d\tau}\right]_{\tau=\tau_{0,n}}\right\}=\mbox{sgn}\left\{ Re\left[\left(\frac{d\lambda}{d\tau}\right)^{-1}\right]_{\tau=\tau_{0,n}}\right\}=\mbox{sgn}\left\{Re\left[\frac{(2\lambda+\delta_a+\delta_b)e^{\lambda\tau}}{\lambda D_{\text{DSNM}}}\right]_{\tau=\tau_{0,n}}\right\}}\\\\
\displaystyle{=\mbox{sgn}\left\{\frac{2\omega_0\cos(\omega_0\tau_{0,n})+(\delta_a+\delta_b)\sin(\omega_0\tau_{0,n})}{\omega_0 D_{\text{DSNM}}}\right\}.}
\end{array}
\]

\begin{figure}
 \centerline{\includegraphics[scale=0.5]{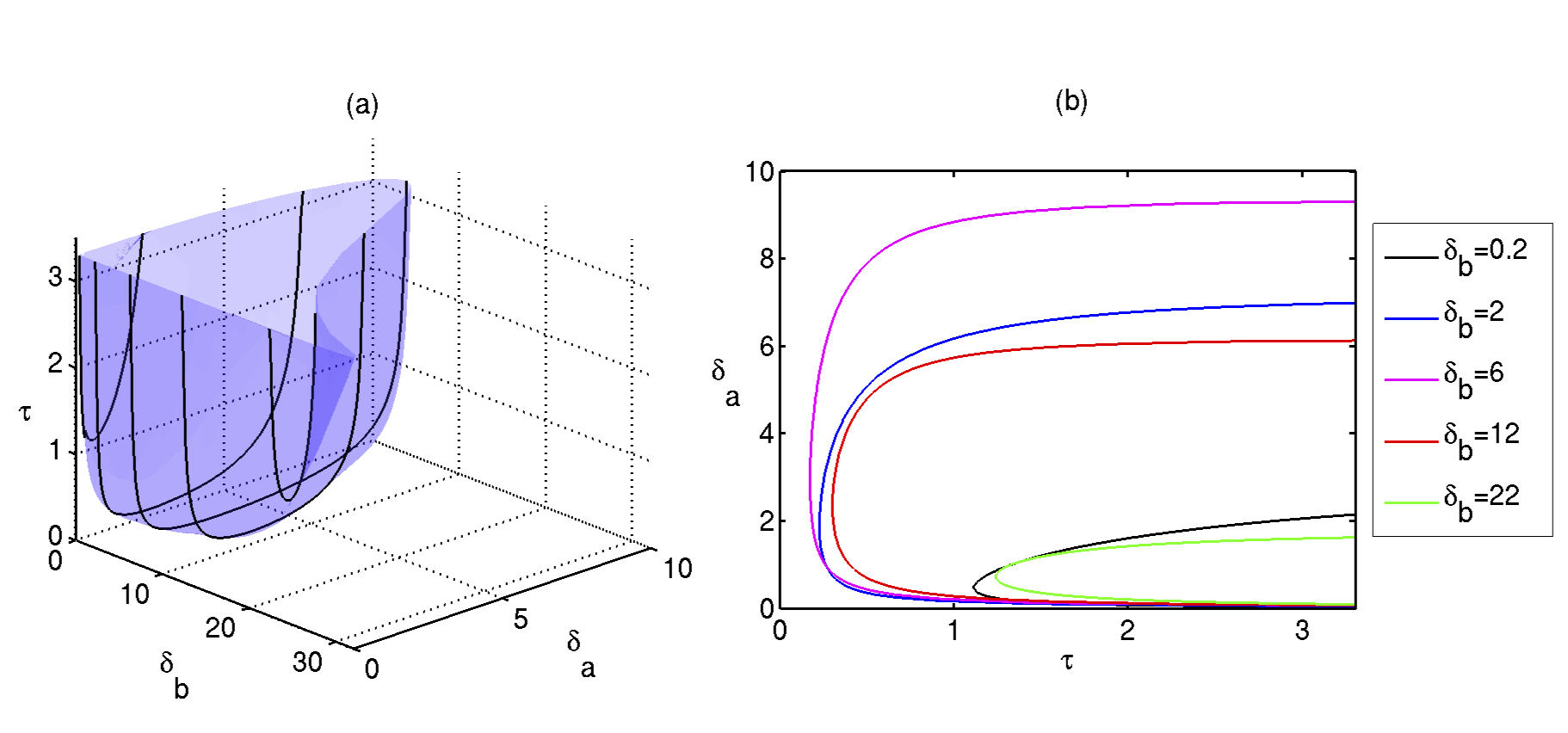}}
 \caption{Stability boundary of the steady state $(\bar{p}_a,\bar{p}_b)$ of the DSNM system (\ref{pab_new}). The steady state is stable below the surface in (a) and to the left of the boundary curves shown in (b). Parameter values are $m_a=m_b=2.35$, $\theta_a=\theta_b=0.21$, $n_a=n_b=3$, and $k_a=k_b=\gamma_a=\gamma_b=1$.\label{fig:Boundary_DSNM_dadb}}
 \end{figure}

\noindent Substituting the expressions for $\cos(\omega_0\tau_{0,n})$ and $\sin(\omega_0\tau_{0,n})$ from the system (\ref{eq:sin_DSNM}) gives
\[
\mbox{sgn}\left\{ \left[\frac{d(Re\lambda)}{d\tau}\right]_{\tau=\tau_{0,n}}\right\}=\mbox{sgn}\left\{\frac{2(\omega_0^2-\delta_a\delta_b)+(\delta_a+\delta_b)^2}{D_{\text{DSNM}}^2}\right\}=\mbox{sgn}\left\{\frac{h'(\omega_0^2)}{D_{\text{DSNM}}^2}\right\}>0.
\]
Hence, the eigenvalues of the characteristic equation cross the imaginary axis at $\tau=\tau_0$ (here, $\tau_0=\tau_{0,0}$) and never cross back for higher values of $\tau$. Thus, we have proved the following result.\\
 
{\bf Theorem 1.} {\it If $\delta_a\delta_b\geq D_{\text{DSNM}}$, the steady state $(\bar{p}_a,\bar{p}_b)$ of the DSNM system (\ref{pab_new}) is stable for all values of the time delay $\tau\geq 0$. If $\delta_a\delta_b<D_{\text{DSNM}}$, this steady state is
stable for $0\leq\tau<\tau_0$, unstable for $\tau>\tau_0$, and undergoes a Hopf bifurcation at $\tau=\tau_0$.}\\

Figure~\ref{fig:Boundary_DSNM_dadb} illustrates the stability boundary of the steady state $(\bar{p}_a,\bar{p}_b)$ of the DSNM system (\ref{pab_new}) depending on the time delay $\tau$ and the protein degradation rates $\delta_a$ and $\delta_b$, with the parameter values taken from Polynikis et al. \cite{polynikis09}. This Figure suggests that for any fixed value of one of such rates, there is only a limited range of positive values of the other degradation rate, for which at a given time delay $\tau$, the positive equilibrium is unstable. For sufficiently high values of $\delta_a$ and $\delta_b$, this steady state is stable regardless of the value of the time delay $\tau$, confirming the result proved in {\bf Theorem 1}.

\begin{figure}
\centerline{\includegraphics[scale=0.45]{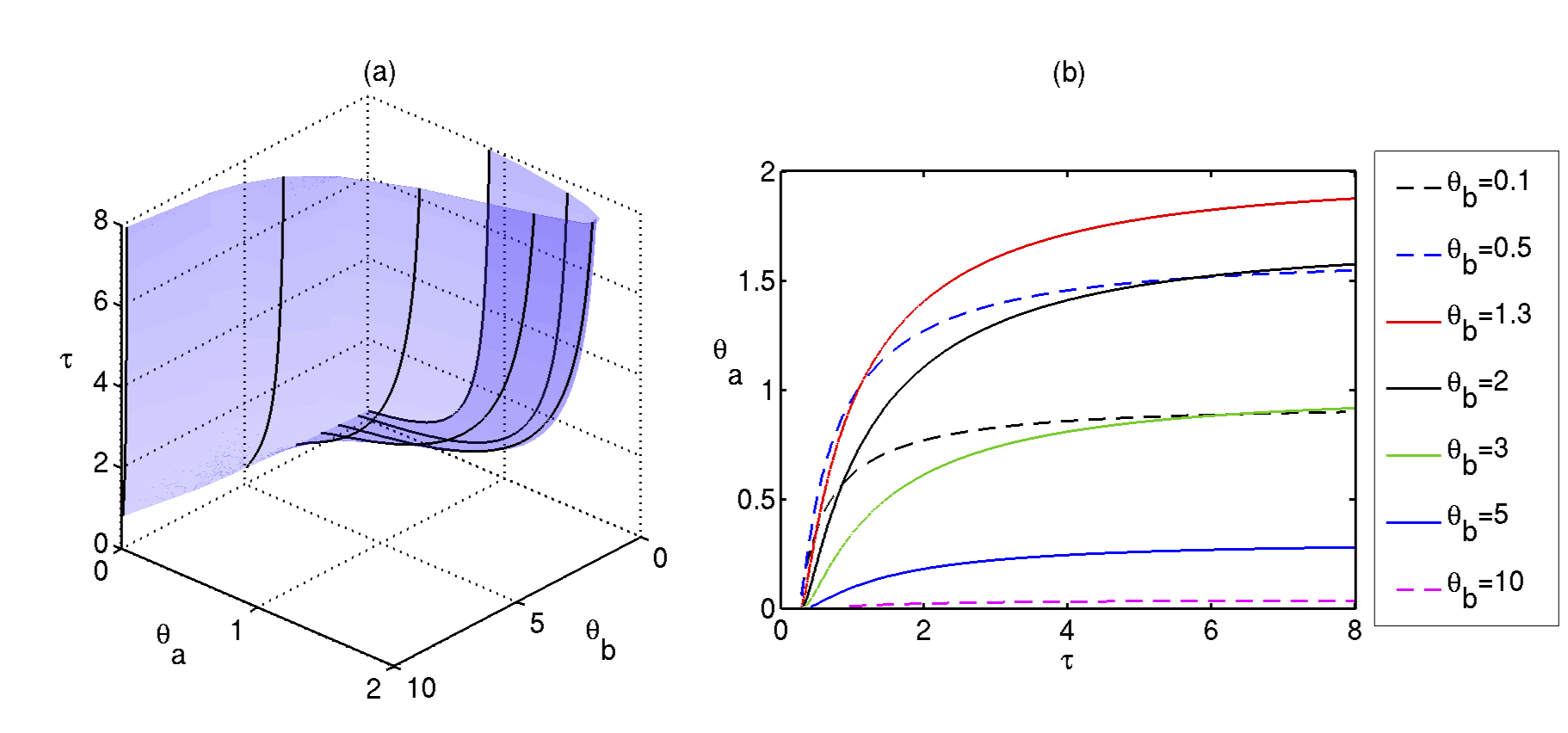}}
\centerline{\includegraphics[scale=0.45]{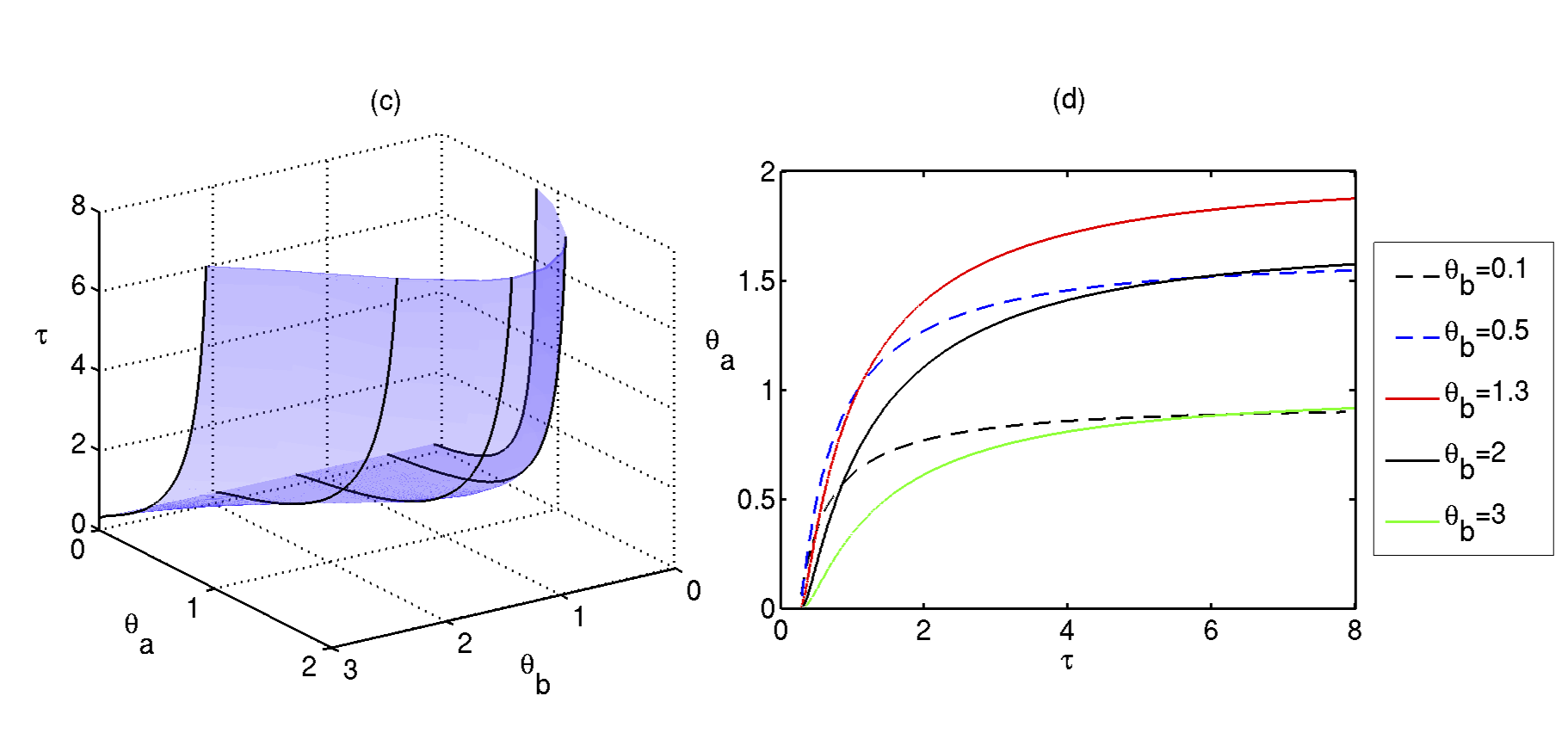}}
\caption{Stability boundary of the steady state $(\bar{p}_a,\bar{p}_b)$ of the DSNM system (\ref{pab_new}). The steady state is stable below the surface in (a), and to the left of the boundary curves shown in (b). Parameter values are $m_a=m_b=2.35$, $n_a=n_b=3$, and $k_a=k_b=\delta_a=\delta_b=\gamma_a=\gamma_b=1$. \label{fig:Boundary_DSNM_thetaathetab}}
\end{figure}

In Fig.~\ref{fig:Boundary_DSNM_thetaathetab} we show how the stability boundary varies depending on the parameters $\theta_a$, $\theta_b$,  and the time delay $\tau$. One observes that for sufficiently high values of the $\theta_b$, the range of possible values of $\theta_a$ for which the steady state is unstable is significantly reduced, thus making the system more prone to support a stable positive equilibrium rather that exhibit oscillations. At the Hopf bifurcation, the associated critical value of the time delay $\tau$ monotonically increases with the parameter $\theta_a$. At the same time, there is a minimum value of the time delay $\tau$, such that for $\tau$ smaller than this value the steady state $(\bar{p}_a,\bar{p}_b)$ is stable for any value of $\theta_a$.

\begin{figure}
\centerline{\includegraphics[scale=0.5]{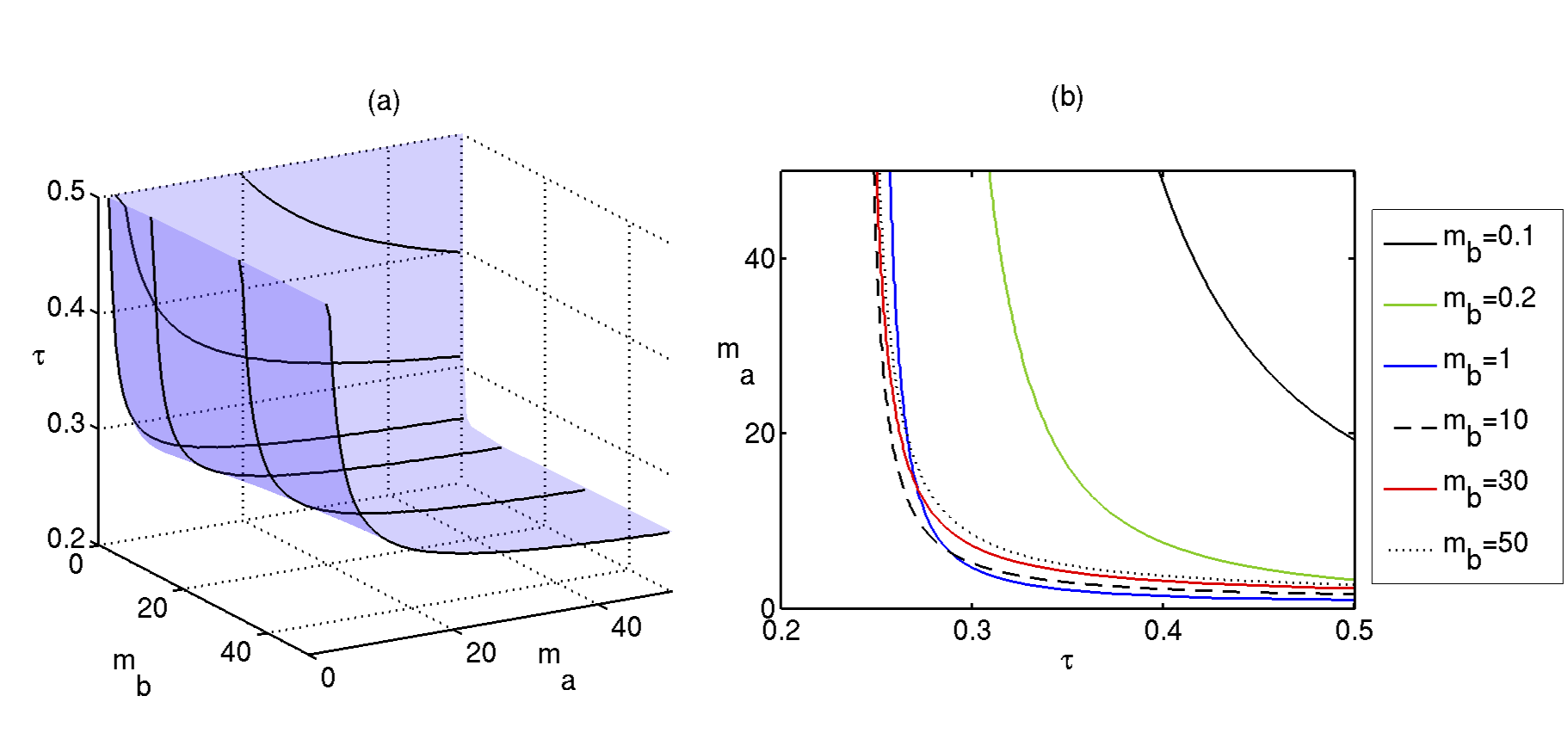}}
 \caption{Stability boundary of the steady state $(\bar{p}_a,\bar{p}_b)$ of the DSNM system (\ref{pab_new}). The steady state is stable below the surface in (a), (c), and to the left of the boundary curves shown in (b), (d). Parameter values are $\theta_a=\theta_b=0.21$, $n_a=n_b=3$, and $k_a=k_b=\delta_a=\delta_b=\gamma_a=\gamma_b=1$. \label{fig:Boundary_DSNM_mamb}}
 \end{figure}

In a similar way, the effects of the transcription rates $m_a$ and $m_b$ are illustrated in Fig.~\ref{fig:Boundary_DSNM_mamb}, which shows that the critical transcription rate of the inhibitor $m_a$ increases with decreasing $\tau$, and similar to Fig.~\ref{fig:Boundary_DSNM_thetaathetab},  below certain value of $\tau$, the steady state $(\bar{p}_a,\bar{p}_b)$ is stable for any value of $m_a$. Qualitatively similar dependence is observed between the critical value of $\tau$ and the transcription rate $m_b$, though this dependence is not completely monotonic.

\begin{figure}
\hspace{-0.5cm}
\centerline{\includegraphics[scale=0.5]{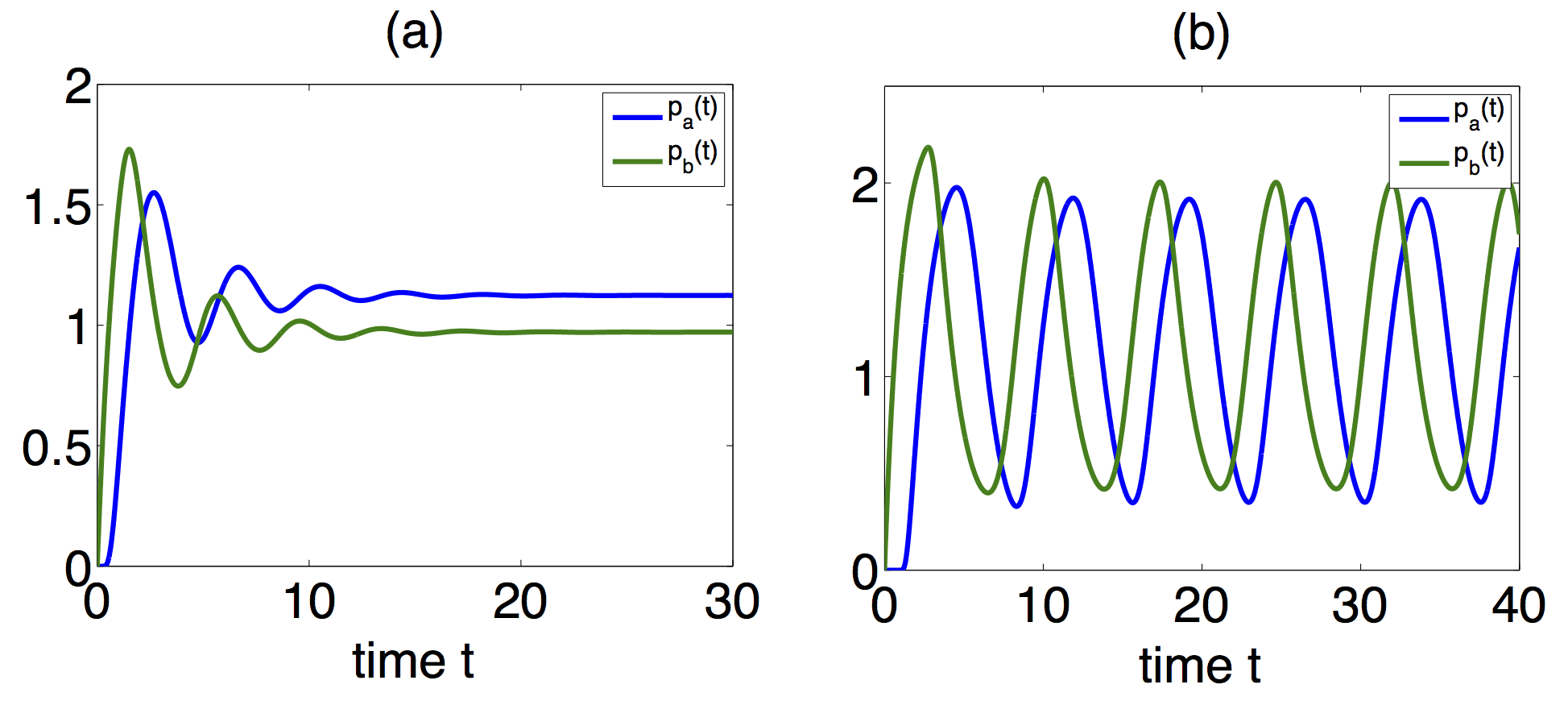}}
 \caption{Numerical solution of the DSNM system (\ref{pab_new}). (a) $\tau=0.5$. (b) $\tau=2$. Parameter values are $m_a=m_b=2.35$, $\theta_a=\theta_b=1$, $n_a=n_b=3$, and $k_a=k_b=\delta_a=\delta_b=\gamma_a=\gamma_b=1$. The critical time delay is $\tau_0=0.9762$.\label{fig:DSNM_solution_thetaathetab}}
 \end{figure}

Figure~\ref{fig:DSNM_solution_thetaathetab} demonstrates how increasing the overall time delay $\tau$ results in a Hopf bifurcation of the steady state $(\bar{p}_a,\bar{p}_b)$ and the emergence of a stable periodic orbit. The shift between individual time series for $p_a$ and $p_b$ can be interpreted in the same way as in predator-prey or activator-inhibitor systems \cite{murray}. In accordance with {\bf Theorem 1}, once the stability of the steady state $(\bar{p}_a,\bar{p}_b)$ is lost, it can never be regained for higher values of $\tau$, so the system will be exhibiting oscillatory behaviour. This result highlights the significance of correct mathematical representation of the transcription and translation processes, since inclusion of transcriptional and translational delays can lead to sustained periodic oscillations even in the simplified model, where such oscillations were impossible when the time delays were neglected.

\section{Analysis of the delayed complete non-linear model (DCNM)}

Linearisation of the full nonlinear DCNM model (\ref{eqn:DCNM}) near the steady state $(\bar{r}_a,\bar{r}_b,\bar{p}_a,\bar{p}_b)$ results in the following characteristic equation
\begin{equation}\label{eqn:char_DCNM}
 (\lambda+\gamma_a)(\lambda+\gamma_b)(\lambda+\delta_a)(\lambda+\delta_b)+D_{\text{DCNM}}e^{-\lambda\tau}=0,
 \end{equation}
 where
 \[
 \displaystyle{D_{\text{DCNM}}=m_am_bk_ak_b\theta_a^{n_a}\theta_b^{n_b}\frac{n_an_b\bar p_a^{(n_a-1)}\bar p_b^{(n_b-1)}}{(\theta_a^{n_a}+\bar p_a^{n_a})^2(\theta_b^{n_b}+\bar p_b^{n_b})^2}=
n_an_b\delta_a\delta_b\frac{\bar{p}^{n_a}_a}{\theta^{n_a}_a+\bar{p}^{n_a}_a}\frac{\theta^{n_b}_b}{\theta^{n_b}_b+\bar{p}^{n_b}_b},}
 \]
 and
 \[
 \tau=\tau_{r_a}+\tau_{r_b}+\tau_{p_a}+\tau_{p_b}.
 \]
It immediately follows from the form of the characteristic equation (\ref{eqn:char_DCNM}) that stability of the steady state $(\bar{r}_a,\bar{r}_b,\bar{p}_a,\bar{p}_b)$ is determined not by individual transcriptional and translational delays, but rather by their overall combined duration. In the case $\tau_{r_a}=\tau_{r_b}=\tau_{p_a}=\tau_{p_b}=0$, the characteristic equation of the DCNM model reduces to the one analysed in Polynikis et al. \cite{polynikis09}.

The characteristic equation (\ref{eqn:char_DCNM}) can be recast in the form
\begin{equation}\label{eqn:char_DCNM_simplified}
\lambda^4+A\lambda^3+B\lambda^2+C\lambda+(D+D_{\text{DCNM}}e^{-\lambda\tau})=0,
\end{equation}
where,
\begin{equation}\label{ABCD_def}
\begin{array}{l}
A=\gamma_a+\gamma_b+\delta_a+\delta_b,\\\\
B=\gamma_a\gamma_b+\gamma_a\delta_a+\gamma_a\delta_b+\gamma_b\delta_a+\gamma_b\delta_b+\delta_a\delta_b,\\\\
C=\gamma_a\gamma_b\delta_a+\gamma_a\gamma_b\delta_b+\gamma_a\delta_a\delta_b+\gamma_b\delta_a\delta_b,\\\\
D=\gamma_a\gamma_b\delta_a\delta_b.
\end{array}
\end{equation}
At $\tau=0$, equation (\ref{eqn:char_DCNM_simplified}) reduces to a quartic
\begin{equation}\label{eqn:char_DCNM_tau0}
\lambda^4+A\lambda^3+B\lambda^2+C\lambda+(D+D_{\text{DCNM}})=0.
\end{equation}
By the Routh-Hurwitz criterion \cite{murray}, the necessary and sufficient conditions for all roots of the equation (\ref{eqn:char_DCNM_tau0}) to have negative real parts are given by:
\[
\begin{array}{l}
\Delta_1 = A >0,\\\\
\Delta_2 = AB-C>0,\\\\
\Delta_3 =ABC-A^2(D+D_{\text{DCNM}})>0,\\\\
\Delta_4 = (D+D_{\text{DCNM}})(ABC-A^2(D+D_{\text{DCNM}})-C^2)=(D+D_{\text{DCNM}})(\Delta_3-C^2)>0.
\end{array}
\]
From the fact that all system parameters are positive, and using the definitions of $A$, $B$ and $C$ in (\ref{ABCD_def}), it follows that
$\Delta_1>0$ and $\Delta_2>0$ for any values of the parameters. Since $D+D_\text{DCNM}>0$, it is sufficient to require $\Delta_4>0$ to ensure that the condition $\Delta_3>0$ is also satisfied. This leads to the following result.\\

\textbf{Lemma 1.} \textit{Let $\tau=0$. The steady state $(\bar{r}_a,\bar{r}_b,\bar{p}_a,\bar{p}_b)$ of the system (\ref{eqn:DCNM}) is stable whenever the condition $ABC-A^2(D+D_{\textnormal{DCNM}})-C^2>0$ holds.}\\

From now on, we will assume that the condition in {\bf Lemma 1} holds, and analyse whether stability can be lost as $\tau$ increases. Since both $D$ and $D_{\text{DCNM}}$ are positive, this means that $\lambda=0$ is not a root of the characteristic equation (\ref{eqn:char_DCNM_simplified}), so once again the stability can only be lost through a possible Hopf bifurcation. To investigate this possibility, we look for solutions of the equation (\ref{eqn:char_DCNM_simplified}) in the form $\lambda=i\omega$ for some real $\omega>0$. Substituting this into the equation (\ref{eqn:char_DCNM_simplified}) and separating into the real and imaginary parts gives
\begin{equation}\label{eqn:cos_DCNM}
\begin{array}{l}
\omega^4-B\omega^2+D=-D_{\text{DCNM}}\cos(\omega\tau),\\\\
-A\omega^3+C\omega=D_{\text{DCNM}}\sin(\omega\tau).
\end{array}
\end{equation}
Squaring and adding these equations yields a quartic equation
\begin{equation}\label{eq:h(z)}
g(z)=z^4+az^3+bz^2+cz+d=0,
\end{equation}
where $z=\omega^2$, and
\[
\begin{array}{l}
a= A^2-2B,\hspace{0.5cm}b= B^2+2D-2AC,\\\\
c= C^2-2BD,\hspace{0.5cm}d= D^2-D_{\text{DCNM}}^2.
\end{array}
\]

Without the loss of generality, suppose that the equation \eqref{eq:h(z)} has four positive real roots, denoted by $z_1,z_2,z_3,z_4$, respectively, which would give four possible values of $\omega$,
\begin{align*}
 \omega_1=\sqrt{z_1},\hspace{2mm} \omega_2=\sqrt{z_2},\hspace{2mm} \omega_3=\sqrt{z_3},\hspace{2mm} \omega_4=\sqrt{z_4}.
\end{align*}
Dividing the two equations in (\ref{eqn:cos_DCNM}) gives
\begin{equation}
 \tan(\omega_k\tau_k)=\frac{A\omega_k^3-C\omega_k}{\omega_k^4-B\omega_k^2+D}\hspace{0.2cm}
\Longrightarrow\hspace{0.2cm} \tau_{k,j} = \frac{1}{\omega_k}\left[\mbox{Arctan}\frac{A\omega_k^3-C\omega_k}{\omega_k^4-B\omega_k^2+D} +j\pi\right],\hspace{0.2cm}k=1,..,4,\hspace{0.2cm}j=0,1,2,...
\end{equation}
Define
\begin{align*}
 \tau_0=\min\limits_{1\leq k\leq 4}\{\tau_{k,0}\}, \hspace{5mm} \omega_0=\omega_{k_0},\hspace{5mm}k_0 \in \{1,2,3,4\},
\end{align*}
then $\tau_0$ is the first value of $\tau>0$ such that the characteristic equation (\ref{eqn:char_DCNM_simplified}) has a pair of purely imaginary roots. We have the following result.\\

\noindent {\bf Theorem 2.} {\it Suppose the conditions of {\bf Lemma 1} hold, and $g'(z_0)>0$, where $g(z)$ is defined in Eq.~(\ref{eq:h(z)}). Then the steady state $(\bar{r}_a,\bar{r}_b,\bar{p}_a,\bar{p}_b)$ of the system (\ref{eqn:DCNM}) is stable for $0\leq\tau<\tau_0$, unstable for $\tau>\tau_0$, and undergoes a Hopf bifurcation at $\tau=\tau_0$.}\\

\noindent {\bf Proof.} The conclusion of {\bf Lemma 1} ensures the steady state $(\bar{r}_a,\bar{r}_b,\bar{p}_a,\bar{p}_b)$ of the system (\ref{eqn:DCNM}) is stable at $\tau=0$, and the fact that the roots of the characteristic equation (\ref{eqn:char_DCNM_simplified}) depend continuously on $\tau$
implies that the steady state $(\bar{r}_a,\bar{r}_b,\bar{p}_a,\bar{p}_b)$ is also stable for sufficiently small positive values of $\tau$.
Since $\tau_0$ is the first positive $\tau$, for which the characteristic eigenvalues lie on the imaginary axis, in order to verify whether or not the steady state actually loses stability at $\tau=\tau_0$, one has to compute the sign of $dRe(\lambda)/d\tau|_{\tau=\tau_0}$. Let $\lambda(\tau)=\mu(\tau)+i\omega(\tau)$ be the root of the characteristic equation (\ref{eqn:char_DCNM_simplified}) near $\tau=\tau_0$, satisfying $\mu(\tau_0)=0$, $\omega(\tau_0)=\omega_0$.
Substituting $\lambda=\lambda(\tau)$ into the Eq.~(\ref{eqn:char_DCNM_simplified}) and differentiating both sides with respect to $\tau$ gives
\[
 \left(\frac{d\lambda}{d\tau}\right)^{-1}=\frac{(4\lambda^3+3A\lambda^2+2B\lambda+C)e^{\lambda\tau}}{\lambda D_{\text{DCNM}}}-\frac{\tau}{\lambda}.
\]
This implies, with $\lambda(\tau_0)=i\omega_0$,
\[
\begin{array}{l}
\displaystyle{\mbox{sgn}\left\{\left[\frac{d(Re\lambda)}{d\tau}\right]_{\tau=\tau_0}\right\}=\mbox{sgn}\left\{ Re \left[\left(\frac{d\lambda}{d\tau}\right)^{-1}\right]_{\tau=\tau_0}\right\}=\mbox{sgn}\left\{ Re \left[ \frac{(4\lambda^3+3A\lambda^2+2B\lambda+C)e^{\lambda\tau}}{\lambda D_{\text{DCNM}}}\right]_{\tau=\tau_0}\right\}}\\\\
\displaystyle{=\mbox{sgn}\left\{\frac{(2B\omega_0-4\omega_0^3)\cos(\omega_0\tau_0)+(C-3A\omega_0^2)\sin(\omega_0\tau_0)}{\omega_0 D_{\text{DCNM}}}\right\}.}
\end{array}
\]
Using the expressions for $\cos(\omega_0\tau_0)$ and $\sin(\omega_0\tau_0)$ from Eq.~(\ref{eqn:cos_DCNM}) gives
\[
\begin{array}{l}
\displaystyle{\mbox{sgn}\left\{\left[\frac{d(Re\lambda)}{d\tau}\right]_{\tau=\tau_0}\right\}=\mbox{sgn}\left\{\frac{4\omega_0^6+(3A^2-6B)\omega_0^4+(2B^2+4D-4AC)\omega_0^2+C^2-2BD}{D_{\text{DCNM}}^2}\right\}}\\\\
\displaystyle{=\mbox{sgn}\left\{\frac{g'(\omega_0^2)}{D_{\text{DCNM}}^2}\right\}>0,}
\end{array}
\]
which means that at $\tau=\tau_0$ a pair of complex conjugate eigenvalues of the characteristic equation (\ref{eqn:char_DCNM_simplified}) crosses the imaginary axis with a positive speed. This implies that the steady state $(\bar{r}_a,\bar{r}_b,\bar{p}_a,\bar{p}_b)$ of the system (\ref{eqn:DCNM}) does lose its stability at $\tau=\tau_0$. $\hfill \square$

\begin{figure}
 \centerline{\includegraphics[scale=0.5]{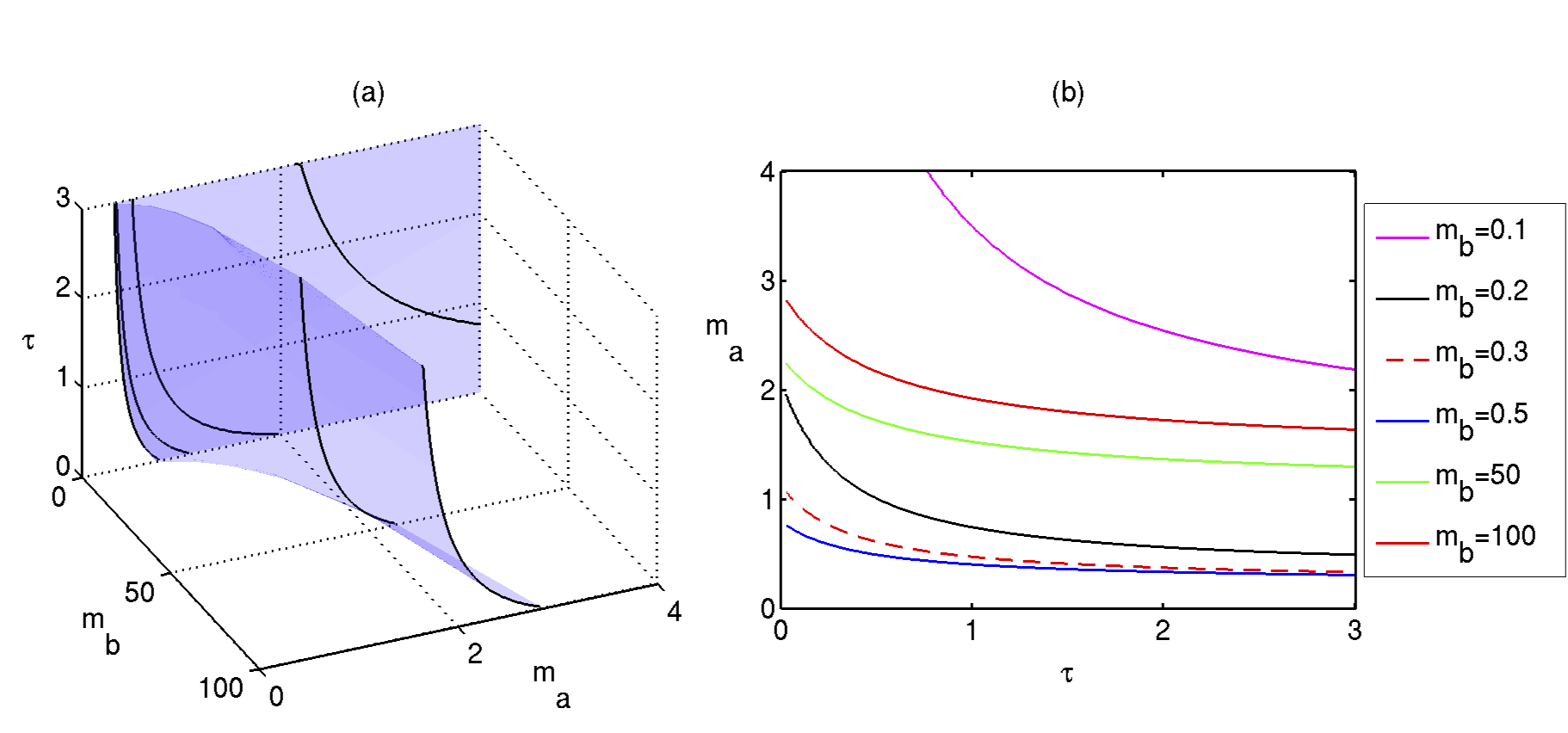}}
 \centerline{\includegraphics[scale=0.5]{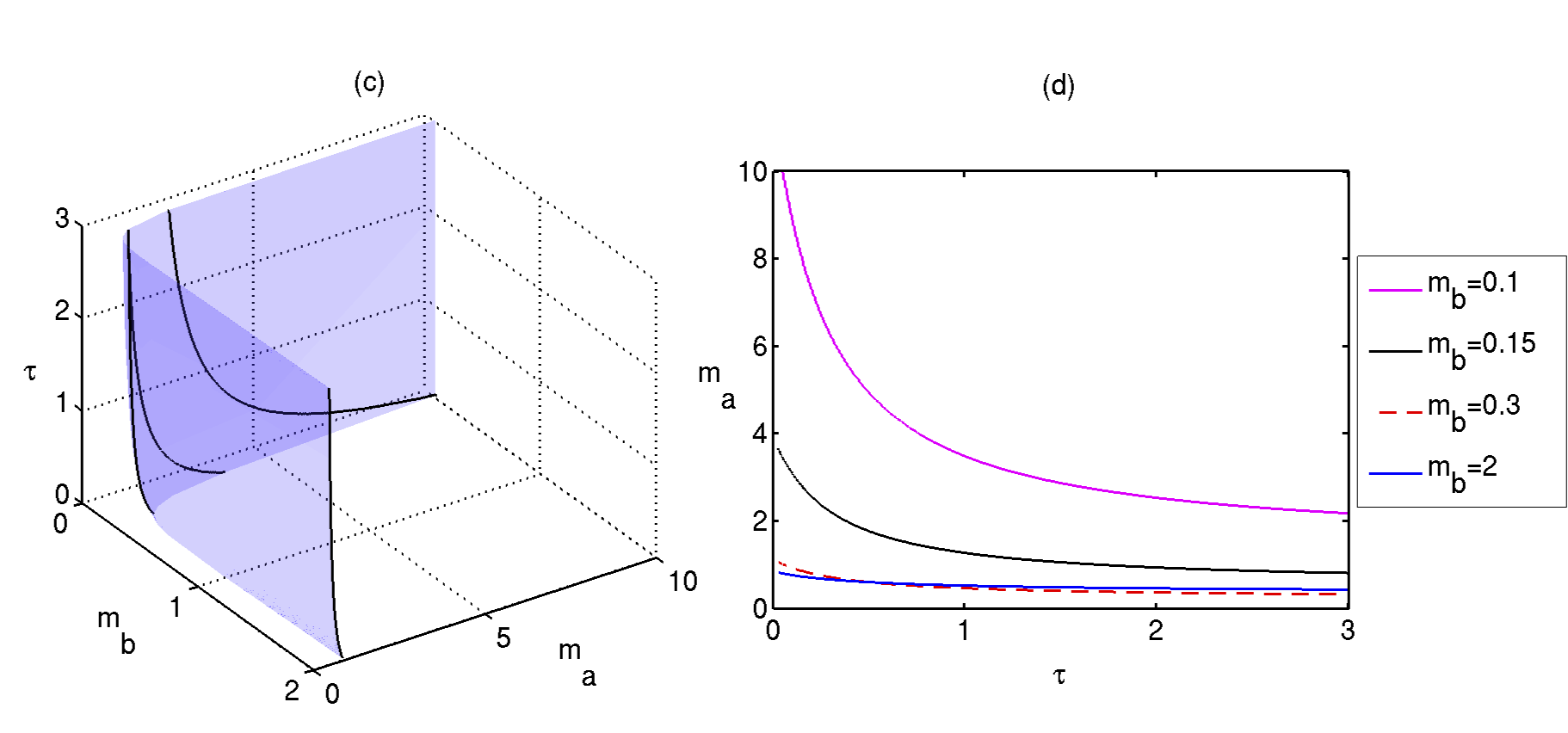}}
\caption{Stability boundary of the steady state $(\bar{r}_a,\bar{r}_b,\bar{p}_a,\bar{p}_b)$ of the DCNM system (\ref{eqn:DCNM}). The steady state is stable below the surface in (a), (c), and below the boundary curves shown in (b), (d). Parameter values: $\theta_a=\theta_b=0.21$, $n_a=n_b=3$ and $k_a=k_b=\delta_a=\delta_b=\gamma_a=\gamma_b=1$. \label{fig:Boundary_DCNM_mamb}}
 \end{figure}

\begin{figure}
\centerline{\includegraphics[scale=0.5]{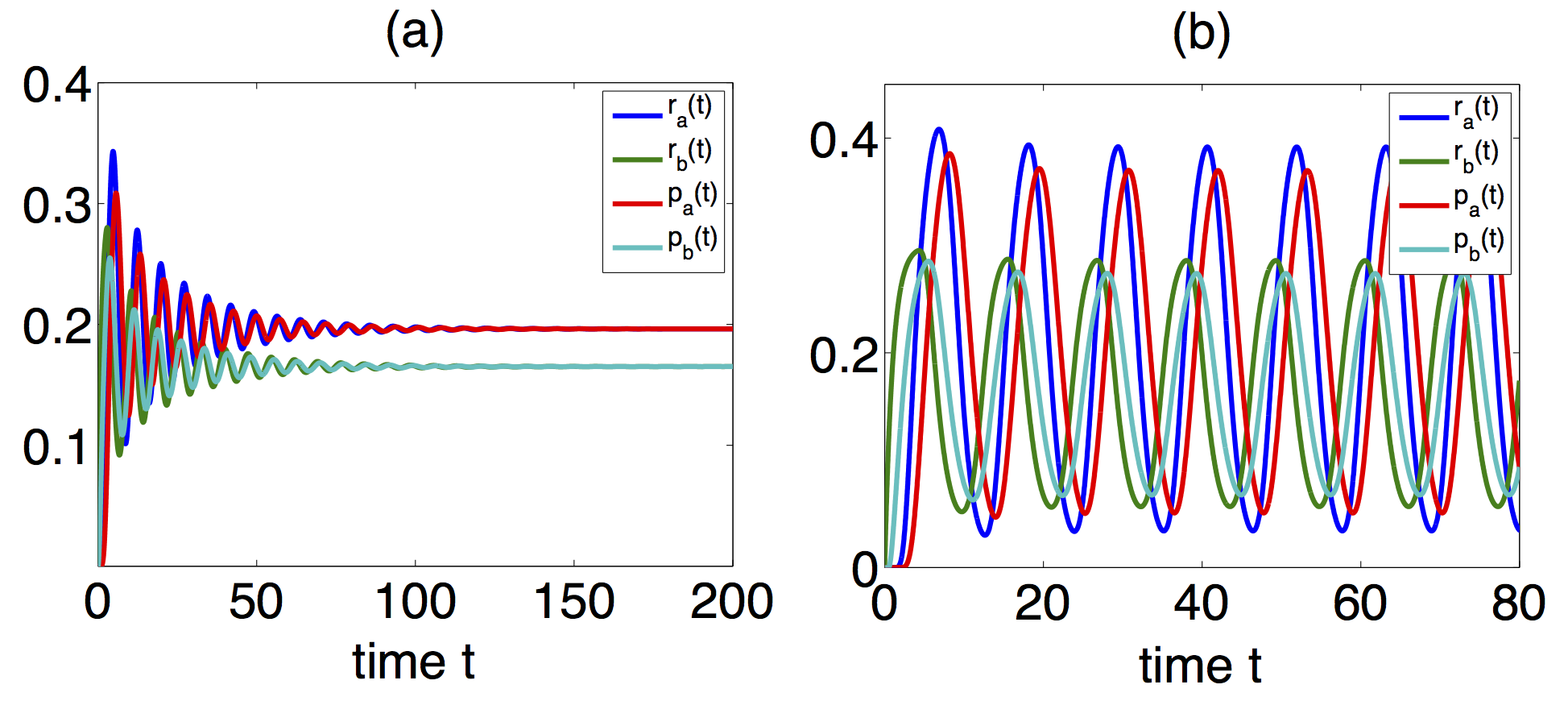}}
\caption{Numerical solution of the DCNM system (\ref{eqn:DCNM}). (a) $\tau=0.25$. (b) $\tau=2$. Parameter values: $m_a=0.6$, $m_b=0.3$, $\theta_a=\theta_b=0.21$, $n_a=n_b=3$, and $k_a=k_b=\delta_a=\delta_b=\gamma_a=\gamma_b=1$. The critical time delay is $\tau_0=0.5314$.\label{fig:DCNM_solution_mamb}}
 \end{figure}

Figure~\ref{fig:Boundary_DCNM_mamb} shows the stability boundary of the steady state $(\bar{r}_a,\bar{r}_b,\bar{p}_a,\bar{p}_b)$ of the system (\ref{eqn:DCNM}) depending on the transcription rates $m_a$ and $m_b$, and the total time delay $\tau$. In a manner similar to that for the simplified model, the critical value of the transcription rate $m_a$ at the Hopf bifurcation reduces with increasing $\tau$. However, a major difference from the DSNM model, as shown in Fig.~\ref{fig:Boundary_DSNM_mamb}, is that now the Hopf bifurcation can take place even at $\tau=0$, as the DCNM system is able to support sustained oscillations \cite{polynikis09}.
In Fig.~\ref{fig:DCNM_solution_mamb} we illustrate the transition from a stable steady state $(\bar{r}_a,\bar{r}_b,\bar{p}_a,\bar{p}_b)$ to a stable periodic solution around this steady state as the time delay passes through the critical value of $\tau=\tau_0$.

\section{Conclusions}

In this review we have discussed various mathematical models for the analysis of GRNs, and focussed on the role played by the transcriptional and translational time delays in the dynamics of a two-gene activator-inhibitor GRN. By reducing the model to the one with a single time delay, we have considered possible behaviour in the quasi-steady state approximation of very fast mRNA dynamics, which has resulted in a lower-dimensional system of DDEs. Due to the presence of time delays, even this simplified model is able to exhibit loss of stability of the positive equilibrium through a Hopf bifurcation, and a subsequent emergence of sustained periodic oscillations, which was not possible in the absence of the time delays, as discussed in Polynikis et al. \cite{polynikis09}. We have found analytically the boundary of the Hopf bifurcation depending on the total time delay and other system parameters, and illustrated different types of behaviour by direct numerical simulations. Our results suggest that once the positive steady state loses its stability, it can never regain it for higher values of the time delay.

We have also studied the stability of a positive steady state in the full system and showed that this steady state can also undergo a Hopf bifurcation depending on the time delay and system parameters. Our analysis extends an earlier result of Polynikis et al. \cite{polynikis09} by showing how the critical values of the parameters at the Hopf boundary change when the time delay increases from zero. Numerical simulations have illustrated the transition from a stable positive steady state to a stable periodic solution as the time delay exceeds its critical value.
 
The work presented in the paper can be extended in several interesting and important research directions. One possibility would be to account for the fact that in most experiments the transcriptional and translational time delay are not fixed but rather obey some form of a delay distribution. Recent work on the effects on delay distribution on system dynamics \cite{blyuss10,kyr11,kyr13} has shown that even for the same mean delay, details of the distribution can also play an important role. He and Cao \cite{HeCao} have used Lyapunov functional approach to derive conditions for global stability of equilibria in some types of GRNs with distributed delays, and it would be insightful to investigate the possibility of extending this methodology to other types of GRNs and various types of delay kernels. Alternatively, one could use the framework of a master stability function for systems with distributed delays \cite{kyr14} to study possible synchronization dynamics in GRNs with a large number of proteins involved.

As it has already been mentioned, in some cases gene expression behaviour is characterised by a switch-like behaviour that can be better modelled using piecewise-linear rather than continuous transcription functions \cite{glass75,casey06}. Whilst some preliminary work has been done recently on the analysis of piecewise-linear systems with discrete time delays, primarily in engineering applications \cite{LS10,sur,sieb}, the dynamics of GRNs with piecewise-linear transcription functions and transcriptional/translational delays has remained completely unexplored. Further inclusion of distributed delays would make such models mathematically very challenging, but it could provide a new level of understanding of GRN dynamics.

Besides providing insights into the dynamics of GRNs, there are several practical ways in which models similar to the one described in this review are helpful for monitoring and treatment of cancer. GRN models based on differential equations coupled with other techniques, such as machine learning and Bayesian networks, have proved effective in identifying specific oncogenes that can be used as biomarkers or drug targets \cite{cheng,ahmad,Li11,folger,raza,zhou12}. Similar kinds of models are useful for modelling cancer cell growth, understanding interactions between tumour growth and immune response, and for analysis of the effects of chemotherapy (or immunotherapy) and drug resistance \cite{cheng,edelman10,banks,tsyg}. The methodology described in this review can be directly used to improve the performance of these models by elucidating the role of transcriptional and translational time delays in GRN dynamics and its impact on various aspects of cancer onset and development.

\section*{Conflict of interest}

The authors declare that there is no conflict of interests regarding the publication of this paper.

\end{document}